\documentclass[conference]{IEEEtran}
\IEEEoverridecommandlockouts
\usepackage{cite}
\usepackage{caption}
\usepackage{amsmath,amssymb,amsfonts}
\usepackage{algorithmic}
\usepackage{graphicx}
\usepackage{textcomp}
\usepackage{xcolor}
\def\BibTeX{{\rm B\kern-.05em{\sc i\kern-.025em b}\kern-.08em
    T\kern-.1667em\lower.7ex\hbox{E}\kern-.125emX}}
\begin{document}

\title{Towards Communication Efficient and Fair Federated Personalized Sequential Recommendation}

\author{\IEEEauthorblockN{Sichun Luo}
\IEEEauthorblockA{\textit{Department of Computer Science} \\
\textit{City University of Hong Kong}\\
Hong Kong, China \\
sichun.luo@my.cityu.edu.hk}
\and
\IEEEauthorblockN{Yuanzhang Xiao}
\IEEEauthorblockA{\textit{Hawaii Advanced Wireless Technologies Institute} \\
\textit{University of Hawaii}\\
Honolulu, HI, USA \\
yxiao8@hawaii.edu}
\and
\IEEEauthorblockN{Yang Liu}
\IEEEauthorblockA{\textit{Institute for AI Industry Research} \\
\textit{Tsinghua University}\\
Beijing, China \\
liuyang.princeton@gmail.com}
\and
\IEEEauthorblockN{Congduan Li}
\IEEEauthorblockA{\textit{School of Electronics and Communication Engineering} \\
\textit{Sun Yat-sen University}\\
Guangzhou, China \\
licongd@mail.sysu.edu.cn}
\and
\IEEEauthorblockN{Linqi Song\IEEEauthorrefmark{1}}
\IEEEcompsocitemizethanks{\IEEEcompsocthanksitem\IEEEauthorrefmark{1}Corresponding author}
\IEEEauthorblockA{\textit{Department of Computer Science} \\
\textit{City University of Hong Kong}\\
Hong Kong, China \\
linqi.song@cityu.edu.hk}
}

\maketitle


\begin{abstract}
Federated recommendations leverage the federated learning (FL) techniques to make privacy-preserving recommendations. Though recent success in the federated recommender system, several vital challenges remain to be addressed: (i) The majority of federated recommendation models only consider the model performance and the privacy-preserving ability, while ignoring the optimization of the communication process; (ii) Most of the federated recommenders are designed for heterogeneous systems, causing unfairness problems during the federation process; (iii) The personalization techniques have been less explored in many federated recommender systems.

In this paper, we propose a \underline{C}ommunication efficient and \underline{F}air personalized \underline{Fed}erated  \underline{S}equential \underline{R}ecommendation algorithm (CF-FedSR) to tackle these challenges. CF-FedSR introduces a communication-efficient scheme that employs adaptive client selection and clustering-based sampling to accelerate the training process.
A fairness-aware model aggregation algorithm that can adaptively capture the data and performance imbalance among different clients to address the unfairness problems is proposed. The personalization module assists clients in making personalized recommendations and boosts the recommendation performance via local fine-tuning and model adaption. Extensive experimental results show the effectiveness and efficiency of our proposed method.\\
\end{abstract}


\section{Introduction}

Sequential recommendations become a crucial task in our daily recommendation scenarios. It is able to learn from a series of correlated browse sequences (e.g., a person may buy shirts, pants, shoes, etc. together), and recommend some new items given the sequence. 
Recent years have witnessed a significant advance and success in sequential recommendation.
The prosperity of neural networks (NN) motivates the community to design NN-based sequential recommendation models \cite{hidasi2015session,kang2018self}.

Recently, user privacy has drawn more concerns, and regulations such as General Data Protection Regulation (GDPR)\footnote{https://gdpr-info.eu/} and California Consumer Privacy Act (CCPA)\footnote{https://oag.ca.gov/privacy/ccpa} are set to protect user privacy. 
To further preserve users' privacy, Federated Learning (FL) techniques \cite{mcmahan2017communication}, as a new privacy-preserving paradigm, are incorporated with the recommender system.
Prior works of federated recommendation tackle the privacy problem by incorporating FL \cite{ammad2019federated,chai2020secure,han2021deeprec,luopaper1}.
Among them, \cite{ammad2019federated} combines the FL with collaborative filtering as a basic paradigm for the federated recommendation. 
Reference \cite{chai2020secure} propose federated matrix factorization for making privacy-preserving recommendations.
Reference \cite{han2021deeprec} resort to tacking the real situation problem by incorporating the sequential recommendation model GRU4REC \cite{hidasi2015session} with FL.

However, current federated recommender systems  are facing some challenges. 
Firstly, the majority of federated recommendation models only consider the model performance and the privacy-preserving ability, while ignoring the optimization of the communication process. Thus the communication overhead could be a heavy burden for the system.
Secondly, most of the federated recommenders are designed for heterogeneous systems, causing unfairness problems during the federation process.
For example, the clients with a large data size would normally share a higher weight in the model aggregation process, then the aggregated model would turn to prefer this client. The model performance in these clients may be obviously better than those of small clients.
Besides, a large proportion of current federated sequential recommender systems do not sufficiently consider personalized recommendations. In other words, all clients share the model with the same parameters. This could degrade the model performance since different clients may have different interests.

To tackle the aforementioned challenges, we hereby propose a \underline{C}ommunication efficient and \underline{F}air  \underline{Fed}erated personalized \underline{S}equential \underline{R}ecommendation algorithm (CF-FedSR), which is a model agnostic algorithm and thus could be fit to other tasks with appropriate adjustments. 
Specifically, we first perform client selection and sampling based on client clustering. The client is clustered according to the latent  representation thus similar clients could benefit from each other for faster training. 
Secondly, we propose a fairness-aware algorithm for model aggregation.
Here we consider the heterogeneity among different clients, and fairness is achieved by designing an adaptive aggregation algorithm to eliminate the impact. Detailedly, we add the fairness coefficient when aggregating the models.
Moreover, the personalized is achieved by local fine-tuning and model adaption.
Our extensive experiments over real-world datasets show the effectiveness of our proposed algorithms, efficiency in time and computing resources, and a tradeoff between privacy and recommendation performance.

Overall, the main contributions of this work are summarized as follows. 

{$\bullet$} We propose a Communication efficient and Fair Federated personalized Sequential Recommendation algorithm. To reduce the communication overhead, 
we utilize adaptive client selection and sampling based on client clustering to accelerate the training process.

{$\bullet$} We investigate the fairness and personalization problem in the federated sequential recommendation, and propose an adaptive
model aggregation and personalization algorithm as the solution. 
The fairness is ensured by adaptively aggregating the uploaded model parameters according to the client feature and performance, and the personalization is achieved by local fine tunes and model adaption.

{$\bullet$} Evaluation on real-world datasets demonstrates the superiority of the proposed model over representative methods. The proposed model brought 9.12\% improvement over competing methods on average while requiring less communication round.

\section{Problem Formulation and Definition}
\label{sec:pf}
In this section, we formally formulate the federated sequential recommendations task and state the current approaches' shortcomings. Also, we define fairness in the federated recommender system.

The federated sequential recommendations task is to train a global recommendation model for a server and many distributed and privacy concerning clients while not disclosing the raw data from the clients.
However, current approaches care less about the communication efficiency, fairness, and personalization of the federated recommenders. We hope to bridge this gap in this paper.
Here we also define the \textit{fairness} of the federated recommender system as the variance between different clients.
For client $\{c_1,...,c_n\}$, with corresponding model performance $\{p_1,...,p_n\}$, then $fairness = variance(\{p_1,...,p_n\})$.

\section{Proposed Method}
In this section, we first describe the overall system architecture, then we detailed introduce our proposed method, including client selection, sampling based on client clustering, fairness-aware model aggregation, and personalization module.

\subsection{Overall System Architecture}
\label{subsec:syss}



\begin{figure}[!t]
\centering
\includegraphics[width=0.47\textwidth,height=0.19\textwidth]{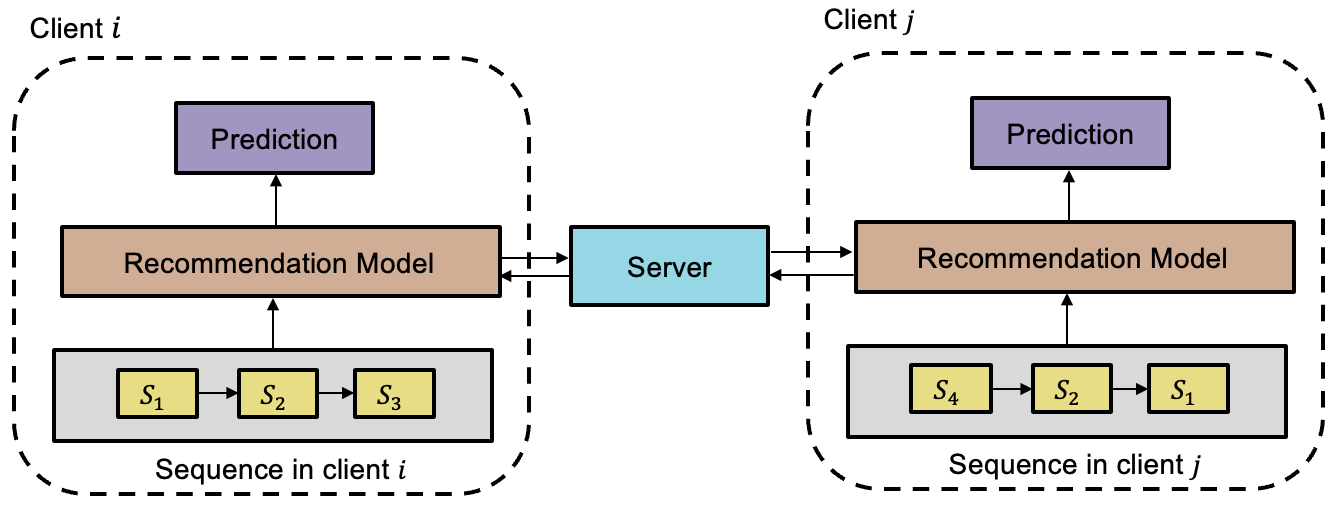}
\caption{Illustration of the overall system architecture}
\label{fig:pathdemo}
\end{figure}

A typical example of federated recommender system is illustrated in Fig.~\ref{fig:pathdemo}.
Here each client has a local recommender engine to recommend items to local users. The local recommender engine performs local model training, exchanges models with the recommendation server, and makes local recommendations. The recommendation server gathers information from all the clients and distributes the aggregated learning model to the clients. In this way, the server and the clients exchange models instead of raw data, which preserves the users' privacy. 

Based on the standard FL algorithm as described beforehand. We try to improve it by designing a communication efficient and fair federated learning algorithm for the sequential recommendation, termed CF-FedSR.

\subsection{Communication efficient and Fair Federated Sequential Recommendation Algorithm (CF-FedSR)}
\label{subsec:pfl}
This section introduces our proposed Communication efficient and Fair Federated Sequential Recommendation algorithm (CF-FedSR).
To enhance the model performance, improve the fairness and reduce the communication overhead  simultaneously, we specifically design client selection, sampling based on client clustering, fairness-aware model aggregation, and a personalization module.
The client selection module introduces the adaptive selector to select appropriate clients to participate in training. 
Then sampling relies on client clustering to adaptively select representative clients.
Also, the parameter aggregation module performs fairness-aware parameter aggregation.
Moreover, the personalization module uses fine-tune techniques for personalized recommendations.

\subsubsection{Client selection}
Federated learning requires lots of transmissions of model parameters between servers and clients. However, we find not all communication is necessary. In the beginning stage, the global model performance is limited due to random initialization.
Allowing all the clients to participate in all training rounds may be suboptimal since small clients may negatively affect the model performance in the beginning stage.
Since for small clients, due to their limited data sizes or low learning rates, they may not contribute enough to improve the model.
Instead, the updates by these small clients may even negatively affect the model aggregation.

To cope with this problem, an adaptive selector is adopted on the server-side to save training time and computing power, as well as to improve the performance.
We introduce two criteria for selecting appropriate clients: (i) the number of data samples $|D_i|$ in client $c_i$ exceeds a threshold $\lambda_1$, or (ii) the training epoch $t_i$ of client $c_i$ exceeds a threshold $\lambda_2$.
If a client $c_i$ satisfy either of the two criteria, we select it to participate in the training.
In this way, the system can get rid of the interference from the updates from small clients in the beginning stage. Through adaptive client selection, the server can train the global model faster and achieve better performance.

\subsubsection{Sampling based on client clustering}
\begin{figure}[t]
\centering
\includegraphics[width=0.44\textwidth,height=0.24\textwidth]{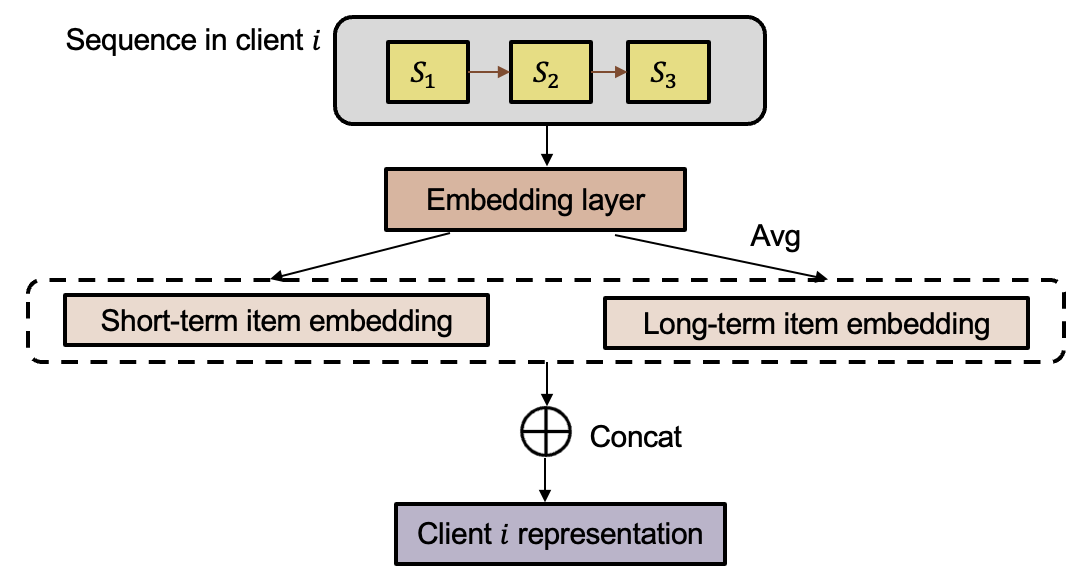}
\caption{Client representation}
\label{fig:client_rep}
\end{figure}

We also propose client clustering to reduce the communication and computation complexity further.
Each client is represented by short-term and long-term interest, as illustrated in Figure~\ref{fig:client_rep}, which is represented by the concatenate of the recent interacted item embeddings.
Specifically, we choose the recent $v_1$ items to represent the user's short-term interest and the recent $v_2$ items to represent the user's long-term interest. Obviously we have $v_2 > v_1$. 
Then we calculate the average embeddings of the selected items. 
After that, we concatenate the long-term and short-term interests to form the client representation.
Several clustering techniques can be adapted to cluster similar users, such as K-means \cite{hartigan1979algorithm} and Mean shift \cite{comaniciu2002mean}. 
We sample clients proportionally from each cluster; thus, the sampled clients should be more representative than the randomly sampled.

\begin{figure}[t]
\centering
\includegraphics[width=0.46\textwidth,height=0.2\textwidth]{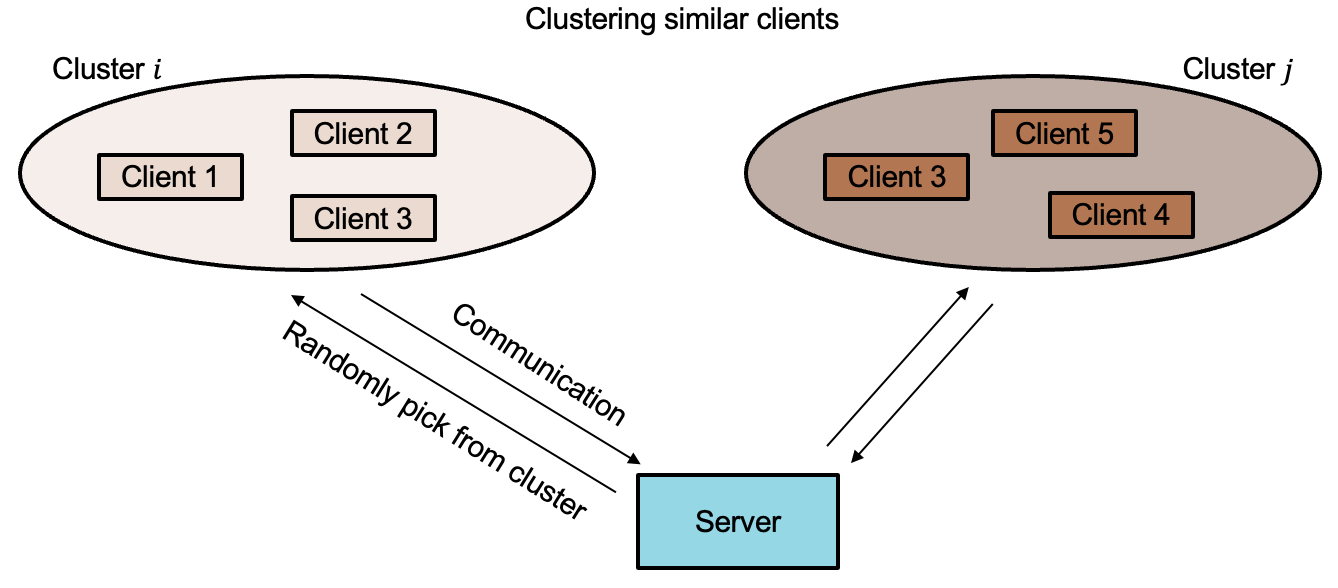}
\caption{Client clustering}
\label{fig:client_clu}
\end{figure}

\subsubsection{\textcolor{black}{Fairness aware parameter aggregation}}

Firstly, we review the FedAvg \cite{konevcny2016federated}, which is a commonly used federation aggregation strategy. The formula is denoted as:
\begin{equation}
w_{t+1} = \sum_{k \in S_t} \frac{n_k}{n} w_k^{t+1}
\end{equation}
where $w_{t+1}$ is the model parameter in time $t+1$, and $n_k$ is the sample size for client $k$.
According to the formula, a larger sample size would result in a more significant contribution to the global model for clients.

However, the FedAvg algorithm suffers from several problems.
Though the overall model performance may be satisfactory, the local model performance for each client may not be promising.
The main reason is that the data for clients are Non-IID, then the local datasets are usually heterogeneous between clients in terms of size and distribution.
Therefore, it is unfair for some clients since they may have a small data sample size and are largely ignored in the parameter aggregation process.

To tackle the heterogeneous problem, we design a weighting strategy to encourage a more fair distribution in this work.
Firstly, we propose a fairness-aware parameter aggregation method to deal with the varying model performance among clients.
To start with, we represent the performance for client $k$ as $p_k$.
For federated recommendation scenarios, commonly adopted evaluation metrics are Hit Rate (HR) and Normalized Discounted Cumulative Gain (NDCG).
Thus we use the sum of evaluation metrics to represent the performance, denoted as:
\begin{equation}
p_{k} =  \text{HR}_k + \text{NDCG}_k,
\end{equation}
Then we normalize it:
\begin{equation}
p_{k}^{\prime} =  \frac{p_k}{\sum_{k \in S_t} p_k}.
\end{equation}
After that, we apply an activation function. The activation function should satisfy decreasing in range $[0,1]$. In this work, the activation function is $f(x) = (\frac{1}{2})^x $.
\begin{equation}
\hat{p}_k = \text{Activation}(p_{k}^{\prime}).
\end{equation}
Then we normalize it again,
\begin{equation}
\hat{p}_k^{\prime} = \frac{p_k}{\sum_{k \in S_t} p_k},
\end{equation}
where the ${p_k^{\prime}}$ is the fairness-aware factor.

Secondly, we alleviate the impact of the data sample size.
For the unfairness caused by the imbalance sample size, one way to eliminate this effect is to reduce the impact of the parameter of data sampling size. In this way, the aggregation formula is denoted as:

We set the weight parameter as $q_k$. Intuitively, it is denoted as:
\begin{equation}
q_k = \frac{n_k}{n}.
\end{equation}
Then we use an activation function. The activation function should satisfies increasing in range $[0,1]$. In this work, the activation function is $f(x) = \sqrt{x} $.
\begin{equation}
\hat{q}_k = \text{Activation}(q_{k}^{\prime}).
\end{equation}
Then we have:
\begin{equation}
\hat{q}_k^{\prime} = \frac{q_k}{\sum_{k \in S_t} q_k}.
\end{equation}

Lastly, we combine these two factors and form the final parameter aggregation function:
\begin{align}
& o_k = \alpha \hat{p}_k^{\prime} + \beta \hat{q}_k^{\prime},\\
& w_{t+1} = \sum_{k \in S_k} \frac{o_k}{\sum_{i \in S_t} o_i} w_k^{t+1},
\end{align}
where $w_{t+1}$ is the global model at time $t+1$, $\alpha$ and $\beta$ are two hyper-parameters.

\subsubsection{Personalization}
Prior researches show that FL could be used to train a high-performing global model for federated recommendation \cite{wu2021fedgnn,qi2020fedrec}. The trained model, however, is more of generality but lacks personalization. Therefore, we fine-tune the global model with the client's local data to make personalized recommendations. We aggregate the global and fine-tuned model to balance generality and personalization. Thus the client could benefit from the personalization process. We denote the process as:
\begin{align}
& w_{local}  = w_{global} - \eta \bigtriangledown L(w, d),\\
& w = \gamma\cdot w_{local} + (1-\gamma)\cdot w_{global},
\end{align}
where $\gamma$ is a hyper-parameter. 
The model could adaptively balance the generalizable and personalization by adjusting the hyper-parameters.
Additionally, note we only transmit the embedding layer in the training process, as the skeleton network, while letting the remaining part of the network be trained localized.


\section{Experiments and Analysis}
\label{sec:exp}
This section presents our experimental results over three real datasets by comparing our proposed algorithms' performance against other reasonable baselines. 
Besides, we demonstrate the effectiveness for each component through an ablation study.
We also analyze the impact of different experimental settings, such as embedding size and different numbers of clusters.
Our experiments are designed to answer the following Research Questions (RQs):

\emph{$\bullet$ RQ1:}  Does CF-FedSR show superiority of existing methods in federated sequential recommendation in terms of model performance, communication cost, and fairness?

\emph{$\bullet$ RQ2:} What is the influence of various components in the CF-FedSR framework? Are those components necessary?

\emph{$\bullet$ RQ3:} What is the impact of the hyper-parameters settings in CF-FedSR?

\subsection{Experiment Setup}
\label{exp_set}

\subsubsection{Dataset Description} 
We have conducted experiments using real-world datasets to evaluate our proposed method.
A detailed description of the total number of user clicks, the number of items and clicks, and the average number of clicks is listed in Table~\ref{tab:description}.
These datasets vary in domains, platforms, and sparsity.
Here we give the detailed description of the datasets.

\emph{$\bullet$ Amazon:} A series of datasets introduced in \cite{mcauley2015image}, comprising large corpora of product reviews crawled from Amazon.com. Top-level product categories on Amazon are treated as separate datasets. We consider two categories, \emph{Beauty} and \emph{Video}. This dataset is notable for its high sparsity and variability.

\emph{$\bullet$ Wikipedia:} This dataset contains one month of edits on Wikipedia \cite{wiki2021}. The editors who made at least 5 edits and the 1,000 most edited pages are filtered out as users and items for recommendation. This dataset contains 157,474 interactions in total.

\begin{table}[t]
\centering
\caption{Description of datasets}
\label{tab:description}
\scalebox{1}
{
\begin{tabular}{lllll}
\hline
Dataset & \# of user & \# of item & \# of click & Avg. length\\
\hline
Beauty & 52204 & 57289 & 394908 & 7.56\\
Video & 31013 & 23715 & 287107 & 9.26\\
Wikipedia & 8227 & 1000 & 157474 & 19.14\\
\hline
\end{tabular}
}
\end{table}

\subsubsection{Baselines}
We aim to explore the performance of federated learning in a deep learning model for sequential recommendations.
We use the \emph{GRU4REC} \cite{hidasi2015session} as the backbone sequential recommendation model. The model combines RNNs to model user sequences for the sequential recommendation.
We compare our proposed CF-FedSR algorithm with the popular \emph{FedAvg} \cite{konevcny2016federated} method.
FedAvg is a classical federated learning algorithm that allows many clients to train a global model collaboratively.
Recall that CF-FedSR is designed to improve FedAvg in communication efficiency, fairness, and personalization.
Also, we compare them with centralized training settings.

\subsubsection{Evaluation Metrics} 
We use the following two metrics to evaluate the performance of our proposed algorithms and existing algorithms.\\
1) \emph{Hit Rate (HR)}. To calculate the proportion of predicted items that are accurate. In this work, we evaluate the top-5 and top-10 items for comparison. \\
2) \emph{Normalized Discounted Cumulative Gain (NDCG)}. To calculate the ranking position of correctly predicted items. When the rank list is predicted correctly, the NDCG is high. If predicted item ranks exceed the recommended limit (5 or 10 in our settings), the NDCG value is zero.

\subsubsection{Experiment Settings} 
For performance evaluation, we use the leave-one-out strategy, following \cite{kang2018self,luopaper2}. For each user, we held out their latest interaction for the test set, the second last for validation and the remaining data for training. 

Also, we adopt a commonly used negative sampling method to reduce heavy computation, in accordance to \cite{kang2018self,luopaper2}.
Specifically, we randomly sample 100 negative items, and rank these items together with the ground-truth item.

\begin{table*}[]
\caption{Experiment results compared with baseline methods. }
\label{overall_exp}
\centering
\scalebox{0.93}[0.93]{
\begin{tabular}{lllllllllllll}
\hline
        & \multicolumn{4}{c}{Beauty}        & \multicolumn{4}{c}{Video}          & \multicolumn{4}{c}{Wikipedia}      \\
Model   & HR@5   & NDCG@5 & HR@10  & NDCG@10 & HR@5   & NDCG@5 & HR@10  & NDCG@10 & HR@5   & NDCG@5 & HR@10  & NDCG@10 \\\hline
Central & 0.3223 & 0.2435  & 0.4482 & 0.2835  & 0.3283 & 0.2648 & 0.4922 & 0.2848  & 0.8982 & 0.8767 & 0.9181 & 0.8828 \\
FedAvg & 0.2651 & 0.1828 & 0.3740 & 0.2173  & 0.2720 & 0.1858 & 0.3987 & 0.2259  & 0.7842    & 0.7606 & 0.8077 & 0.7681  \\
CF-FedSR  & 0.2869 & 0.2036 & 0.4073 & 0.2509  & 0.2951 & 0.2008 & 0.4223 & 0.2517  & 0.8469    & 0.8199 & 0.8722 & 0.8279  \\
Impro.  & 8.22\% & 11.38\% & 8.90\% & 15.46\% & 8.49\% & 8.07\% & 5.92\% & 11.42\% & 8.00\% & 7.80\% & 7.99\% & 7.79\%\\\hline
\end{tabular}}
\end{table*}

\subsubsection{Implementation Details} 
The number of users used in each round of model training is 128, and the total number of the epoch is 200.
The ratio of dropout is 0.3. Adam \cite{kingma2014adam} is selected as the optimizer,
and the default learning rate is 0.001. 
We tune hyper-parameters using the validation set, and
terminate training if validation performance doesn't improve
for five successive epochs.
Unless stated otherwise, we use the same hyper-parameters of FedAvg and CF-FedSR.
We report the average performance scores over the five repetitions.

Our evaluation target is to compare the recommendation performance in a federated context,
thus we don't compare it to other centralized algorithms since they're not well suited to federated settings.

\subsection{ Performance Comparison (RQ1) }

\subsubsection{Evaluation on model performance}
We have compared different methods' performances in Table~\ref{overall_exp}. 
We have the following observations:

$\bullet$ 
Our proposed CF-FedSR significantly outperforms the state-of-the-art federated recommender systems in all datasets.
Compared with FedAvg, CF-FedSR achieves on average 7.92\% and 10.32\% relative improvements on HR and NDCG, respectively. 
Several advantages of CF-FedSR support its superiority: 
(1)  client selection and sampling helps to find more representative and useful clients to participate in training;
(2) fairness aware aggregation algorithm help to bridge the gap between clients with various data distribution and facilitate the training.

$\bullet$ The centralized models perform better than those decentralized models. 
Compared with CF-FedSR, centralized models achieves in average 10.25\% and 15.20\% relative improvements in HR and NDCG, respectively. 
There are two reasons: 
(1) centralized models models are better as they can directly train the model without noise or pseudo-labeled items. 
In the meantime, the federated learning framework achieves better privacy-preserving ability than centralized training because clients do not reveal raw data to the server.
It could be viewed as the tradeoff of the recommendation performance to privacy.
(2)
though FL theoretically would have similar performance with the centralized settings. However, due to the data heterogeneous, distributed systems are harder to train since they lack the capacity to model the global data structures.

%

\begin{table}[]
\caption{Comparison of overall performance and converge speed among CF-FedSR and competing methods.}
\label{converge}
\centering
\begin{tabular}{lllll}
\hline
Model  & HR@10   & NDCG@10 & Converge round \\\hline

FedAvg &    0.3740      &   0.2173      &    \~75               \\  
CF-FedSR  & 0.4073  & 0.2509  &  \~67           \\
Impro.  & 8.90\%  & 15.46\%  &  -10.67\%           \\\hline
\end{tabular}
\end{table}

\subsubsection{Evaluation on communication efficiency}
We evaluate the converge speed together with the model performance in Table~\ref{converge}.
We observe that the CF-FedSR converges much more quickly than FedAvg.
Thus the communication cost is saved by transmitting less round.
The reason is that we adopt several techniques to reduce communication loss.
We strictly limit the participation of small clients at the beginning stage of training, which prevents the inference from these small clients.
Additionally, we carefully select representative clients to participate through clustering, making the system converge faster.
In a nutshell, CF-FedSR can enhance the model performance while boosting communication efficiency.


\subsubsection{Evaluation on fairness}
\begin{figure}[t]
\centering
\caption{Comparison of overall performance and fairness among CF-FedSR and competing methods.}
\label{fairness}
\centering
\includegraphics[width=0.46\textwidth,height=0.17\textwidth]{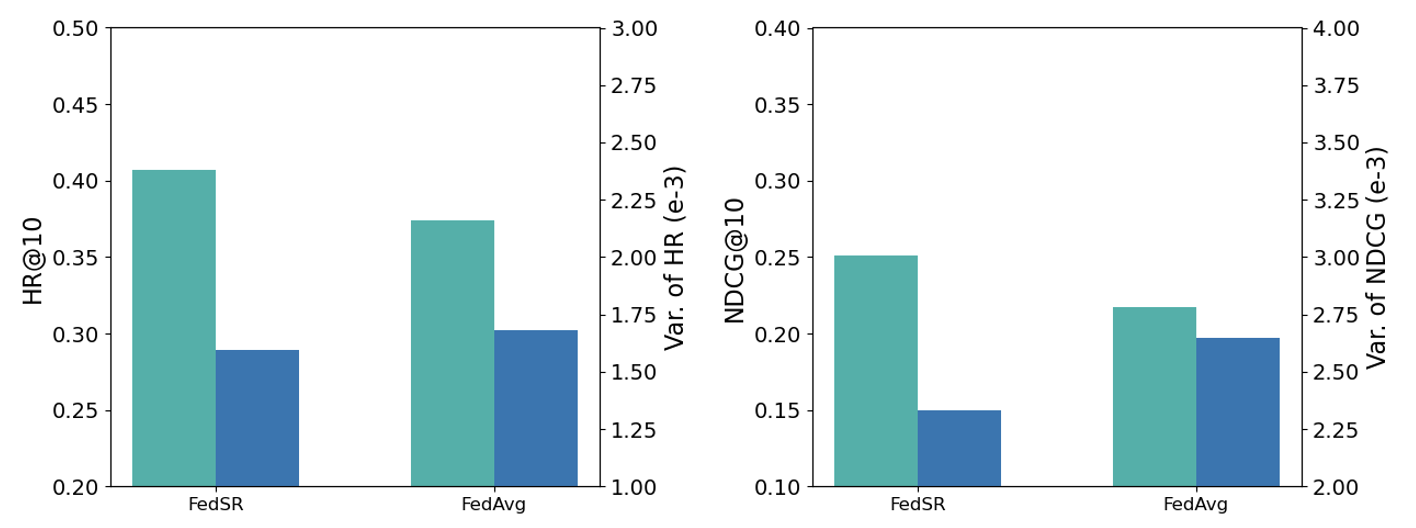}
\end{figure}

Recall that we define the \textit{fairness} as the variance among clients in Section~\ref{sec:pf}. A lower value of variance means the system is more fair.
The experimental results is shown in Figure~\ref{fairness}.
The result demonstrate that our proposed method could decrease the variance and boost the model performance in the meanwhile, showing the effectiveness of our proposed method.
The superiority of our proposed method come from our specially designed adaptive aggregation algorithm.

\subsection{Ablation Study (RQ2)}

Specifically, we repeat the experiment by removing one module from the proposed CF-FedSR model and test the performance.
The detailed information of variants is listed as follows.

\textbf{$\bullet$ CF-FedSR-Variation 1 (w/o client selection \& sampling)}: The proposed CF-FedSR without the client selection \& sampling component.

\textbf{$\bullet$ CF-FedSR-Variation 2 (w/o fair aggregation)}: The proposed CF-FedSR without fair aggregation component.

\textbf{$\bullet$ CF-FedSR-Variation 3 (w/o personalization)}: The proposed CF-FedSR without the personalization module.

From Table~\ref{ablation}, we observe that the CF-FedSR achieves the best results on all datasets, which verifies the importance of each critical module.
Among them, the experimental results of CF-FedSR-Variation 3 drop sharply, which proves that the lack of personalization could significantly decrease the learning ability of the framework.

\begin{table}[!h]
\caption{Ablation study.}
\label{ablation}
\centering
\scalebox{0.98}[0.98]{
\begin{tabular}{lllll}
\hline
      & \multicolumn{4}{c}{Wikipedia}      \\
  Model           & HR@5   & NDCG@5 & HR@10  & NDCG@10 \\\hline
CF-FedSR             & 0.8469 & 0.8199 & 0.8722 & 0.8279  \\
CF-FedSR-Variation 1   & 0.8039 & 0.7631 & 0.8396 & 0.7751  \\
CF-FedSR-Variation 2     & 0.8259 & 0.7968 & 0.8610 & 0.8083  \\
CF-FedSR-Variation 3 & 0.7984 & 0.7703 & 0.8358 & 0.7801 \\\hline
\end{tabular}
}
\end{table}

\begin{table}[!h]
\caption{CF-FedSR performance with respect to different embedding sizes $d$.}
\label{para_d}
\centering
\begin{tabular}{lllll}
\hline
Dataset & \multicolumn{4}{c}{Beauty}         \\\hline
Dimension        & HR@5   & NDCG@5 & HR@10  & NDCG@10 \\\hline
d=10    & 0.2631 & 0.1986 & 0.3689 & 0.2296  \\
d=20    & 0.2726 & 0.2003 & 0.3821 & 0.2389  \\
d=50    & 0.2869 & 0.2036 & 0.4073 & 0.2509  \\
d=100   & 0.2965 & 0.2073 & 0.4124 & 0.2516 \\\hline
\end{tabular}
\end{table}

\begin{table}[!h]
\caption{CF-FedSR performance with respect to different cluster number $k$.}
\label{para_k}
\centering
\begin{tabular}{lllll}
\hline
Dataset     & \multicolumn{4}{c}{Beauty}        \\\hline
\# clusters & HR@5   & NDCG@5 & HR@10  & NDCG@10 \\\hline
k=3         & 0.2846 & 0.2013 & 0.4029 & 0.2453  \\
k=5         & 0.2869 & 0.2036 & 0.4073 & 0.2509  \\
k=10        & 0.2912 & 0.2065 & 0.4068 & 0.2527 \\\hline
\end{tabular}
\end{table}

Without the client selection \& sampling module, the performance of the framework still decreases obviously. 
Throughout these three ablation experiments, it turns out that each module improves the model performance from different aspects and is meaningful.

\subsection{Hyper-parameter study (RQ3)}
\subsubsection{Sensitivity of CF-FedSR to embedding size}
Table~\ref{para_d} illustrates the analysis of the effect of embedding on the \emph{Beauty} dataset. 
We repeat the experiment with default parameters and with different values for the number of embedding size, i.e. with $d \in \{10, 20, 50, 100\}$.
We find that performance is continually improved with the increase of embedding size.
Presumably, this is because higher dimension of item embedding could bring more information from different aspect.
However, a larger embedding size would have a higher communication cost. Thus it is a tradeoff between the communication cost and the model performance.

\subsubsection{Sensitivity of CF-FedSR to number of clusters}

This experiment demonstrates the sensitivity of CF-FedSR to the number of clusters hyper-parameter. We ran CF-FedSR using the default parameters in subsection 5.1 but with different values for the number of clusters, i.e. with $k \in \{3, 5, 10\}$. The result of this experiment is presented in Table~\ref{para_k}.
We observe that its performance improves and then drops when the number of cluster increases from
3 to 10.

\section{Conclusion}
\label{sec:con}
In this paper, we study the problem of the federated sequential recommendations and propose the CF-FedSR framework.
The key component for CF-FedSR includes client selection, sampling based on client clustering, fairness-aware model aggregation, and a personalization module.
These techniques are adopted to ensure fairness and personalization in the federated recommendation, boost the recommendation and ease the communication cost in the meanwhile.
We conducted extensive experiments over real datasets and observed that our proposed framework effectively makes federated sequential recommender systems more accurate and reduces the communication overhead.

\section{Acknowledgements}
This work was supported in part by the Changsha Science and Technology Program International and Regional Science and Technology Cooperation Project under Grants kh2201026, the Hong Kong RGC grant ECS 21212419, the Technological Breakthrough Project of Science, Technology and Innovation Commission of Shenzhen Municipality under Grants JSGG20201102162000001, InnoHK initiative, the Government of the HKSAR, Laboratory for AI-Powered Financial Technologies, the Hong Kong UGC Special Virtual Teaching and Learning (VTL) Grant 6430300, and the Tencent AI Lab Rhino-Bird Gift Fund.

\bibliography{refs}
\bibliographystyle{IEEEtran}








\end{document}